\documentclass[useAMS,usenatbib,a4]{mn2e}
\usepackage[dv ips]{graphicx}
\usepackage{multirow}
\usepackage{natbib}
\usepackage{upgreek,amssymb}

\newcommand{\mnras}{MNRAS}
\newcommand{\apj}{ApJ}

\newcommand{\apjl}{ApJL}
\newcommand{\pasp}{PASP}
\newcommand{\aap}{A\&A}

\newcommand{\msun}{\mbox{M$_{\odot}$}}

\title[Binary progenitors for supernova iPTF13bvn]{Possible binary progenitors
  for the type Ib supernova iPTF13bvn}
\author[J. J. Eldridge et al.]{J. J. Eldridge$^{1}$ \thanks{E-mail:
    j.eldridge@auckland.ac.nz}, Morgan Fraser$^{2}$, Justyn
  R. Maund$^{3,4}$, Stephen J. Smartt$^{3}$ \\ $^{1}$Department of
  Physics, University of Auckland, Private Bag 92019, Auckland, New
  Zealand\\ $^{2}$Institute of Astronomy, University of Cambridge,
  Madingley Road, Cambridge CB3 0HA, UK\\ $^{3}$Astrophysics Research
  Center, School of Mathematics and Physics, Queen's University
  Belfast, Belfast BT7 1NN, UK\\ $^{4}$Department of Physics \&
  Astronomy, University of Sheffield, Hicks Building, Hounsfield Road,
  Sheffield S3 7RH, UK }

\pagerange{\pageref{firstpage}--\pageref{lastpage}} \pubyear{2005}
\begin{document}
\maketitle
\label{firstpage}

\begin{abstract}
\cite{Cao13} reported a possible progenitor detection for the type Ib
supernovae iPTF13bvn for the first time. We find that the progenitor
is in fact brighter than the magnitudes previously reported by
approximately 0.7 to 0.2 mag with a larger error in the bluer
filters. We compare our new magnitudes to our large set of binary
evolution models and find that many binary models with initial masses
in the range of $10$ to $20M_{\odot}$ match this new photometry and
other constraints suggested from analysing the supernova. In addition
these lower mass stars retain more helium at the end of the model
evolution indicating that they are likely to be observed as type Ib
supernovae rather than their more massive, Wolf-Rayet counter parts. We are
able to rule out typical Wolf-Rayet models as the progenitor because
their ejecta masses are too high and they do not fit the observed SED
unless they have a massive companion which is the observed source at
the supernova location. Therefore only late-time observations of the
location will truly confirm if the progenitor was a helium giant and
not a Wolf-Rayet star.
\end{abstract}

\begin{keywords}
stars: evolution -- binaries: general -- supernovae:general -- supernovae: iPTF13bvn
\end{keywords}

\section{Introduction}

Massive stars end their lives in the explosive death throes of a
core-collapse supernova (SNe). These SNe are classified according to
their observed spectra and lightcurves; in the first instance by the
presence or absence of hydrogen in the SN spectrum -- hydrogen rich
SNe are classified as ``Type II'', while hydrogen-deficient SNe are
``Type I''. Type I SNe are further divided\footnote{We note that Type
  Ia SNe are hydrogen-deficient supernova from a thermonuclear
  explosion mechanism in a carbon-oxygen white dwarf which we do not
  consider here.} into Types Ib and Ic (collectively termed Type Ibc),
which are helium rich and helium poor respectively. While the
progenitors of Type II SNe have been directly identified in
pre-explosion images as H-rich supergiants between 8 and 16
\msun\ \citep[and references therein]{Sma09}, the progenitors of Type
Ibc SNe have remained elusive.

The two likely candidates for the progenitors of Type Ibc SNe are
single, massive Wolf-Rayet (WR) stars \citep{Gas86}, or lower mass
stars in binaries \citep{Pod92}. In both cases, the progenitor will be
stripped of its H and/or He envelope. \cite{Eld13} presented a sample
of nearby Type Ibc SNe with pre-explosion {\it Hubble Space Telescope}
({\it HST}) imaging, but found no progenitor
candidates. \citeauthor{Eld13} suggested that this was evidence that a
number of the progenitors of these supernovae must been the result of
an interacting binary star, as previously suggested from the relative
rates of different SN types \citep{DV98,Eld08,Smi11}. There is also
growing additional evidence from statistical samples of ejecta masses
that most Ib/c SNe are from low mass stars in binaries
\citep{drout,lyman,bianco}

Last year \cite{Cao13} presented the detection of a possible
progenitor candidate in {\it HST} images for the Type Ib supernova
iPTF13bvn in the nearby galaxy NGC 5806. From the magnitudes they
report for the progenitor candidate, along with indirect constraints
on its radius and mass-loss rate from observations of the SN itself,
they suggested the progenitor of iPTF13bvn was consistent with a
single WR star. \cite{Gro13} compared their single-star models to the
constraints and found a possible initial mass range between 31 to 35
\msun\ for the progenitor. However, follow-up observations of
iPTF13bvn yielded a bolometic lightcurve which, when fitted with a
hydrodynamic model for the SN ejecta, implied a pre-explosion mass of
$\sim$3.5 \msun\ \citep{Fre14,Ber14}. Such a low mass is inconsistent
with a single WR star, which in the models of \citeauthor{Gro13} would
have a pre-explosion mass of $\sim$11 \msun. \citeauthor{Ber14}
further presented modelling of a binary progenitor system consisting
of a 19 \msun\ primary and a 20 \msun\ secondary which could match the
pre-explosion constraints from {\it HST}. It is important to note that
\citet{yoon} predicted it would be easier to observe such a low-mass
helium star than a Wolf-Rayet star as the progenitor of a type Ib/c
SN.

In this letter we first reanalyse the pre-explosion images of the site
of the SN. We use late-time {\it HST} images to revisit the astrometry
and photometry of the progenitor candidate. We then compare the
derived observational constraints for the progenitor of iPTF13bvn to
our grid of binary evolution models from the BPASS (Binary Population
and Spectral Synthesis, \texttt{http://bpass.auckland.ac.nz})
code. While \cite{Ber14} have presented a binary progenitor scenario
for iPTF13bvn, they note that their solution is not
unique. Furthermore, we find the photometry of \cite{Cao13} to which
\cite{Ber14} fit their models to underestimates the progenitor
candidates magnitudes. With our grid of models, we can compare a wide
range of binary systems to the progenitor of iPTF13bvn, and constrain
the allowed parameter space of the system.

In the following, we adopt a distance of 22.5$\pm$2.4 Mpc,
$\upmu=31.76\pm0.36$ mag
towards NGC 5806, as used by \citeauthor{Cao13} and \citeauthor{Fre14}
from \cite{Tully2009}. While the foreground reddening towards NGC 5806
is low \mbox{(E(B-V)=0.045)} mag from the \cite{Sch11} dust maps, the
host galaxy reddenings adopted by \citeauthor{Ber14},
\citeauthor{Cao13} and \citeauthor{Fre14} differ
significantly. \citeauthor{Ber14} find, E(B-V)=0.17$\pm$0.03 mag,
under the assumption that the colour evolution of iPTF13bvn matches
that of other Type Ibc SNe, however \citeauthor{Cao13} adopt a much
lower value of, E(B-V)=0.03 mag, from the strength of the Na D lines
in high resolution spectra. In this paper, we consider both possible
values and find progenitors systems that fit between these two
extinctions to take account of the uncertainty in the amount of dust.

\section{On the progenitor detection}
\label{s2}
\cite{Cao13} identified a progenitor candidate for iPTF13bvn in
pre-explosion {\it HST} Advanced Camera for Surveys (ACS) images,
acquired as part of program GO-10187 (PI: Smartt). These observations
were acquired with the Wide Field Channel (WFC; pixel scale
0.05\arcsec~pix$^{-1}$) of ACS on 2005 March 10 using the $F435W$
(1600s), $F555W$ (1400s) and $F814W$ (1700s) filters. A key
outstanding question in the \citeauthor{Cao13} analysis, however, was the
level of agreement between the position of the progenitor candidate on
the pre-explosion image and the transformed SN position derived from
post-explosion adaptive optics images. \cite{Fre14} presented a
re-analysis of the position of iPTF13bvn using {\it HST}+WFC3
observations of iPTF13bvn, and found the \citeauthor{Cao13} progenitor
candidate to be coincident with the SN. Using the same data as
\citeauthor{Fre14}, we have performed an independent analysis of the
position of iPTF13bvn.

New {\it HST} Wide Field Camera 3 (WFC3; pixel scale
0.04\arcsec~pix$^{-1}$) Ultraviolet and Visual (UVIS) imager
observations ($F555W$ 1200s) were taken of iPTF13bvn on 2013 September
2 (program GO-12888, PI Van Dyk). Using the {\it astrodrizzle} task
within the {\it drizzlepac} package, the under sampled WFC3 {\texttt
  \_flt} images were drizzled \citep{Fru02} onto a finer pixel scale,
yielding a distortion corrected combined image with a pixel scale of
0.025\arcsec. The pre-explosion ACS images were taken at the same
pointing, and so drizzling could not be used to improve their spatial
resolution. However, the two individual {\texttt \_flc} frames were
still combined with {\it astrodrizzle} (although with an output pixel
scale of 0.05\arcsec) to remove cosmic rays and correct for the
geometric distortion of ACS.

Using 29 point sources identified in both the ACS {\it F555W} and WFC3
frames, we derived a geometric transformation between the pre- and
post-explosion images with an RMS error of 0.38 ACS pixels (19
mas). The pixel coordinates of iPTF13bvn were then measured on the
post-explosion WFC3 image (as the SN is bright, the uncertainty on its
position is negligible in all of the following) and transformed to the
ACS frame. We find the progenitor candidate of \cite{Cao13} to
be offset by only 7 mas from the transformed position of iPTF13bvn,
and hence formally coincident, as also found by \cite{Fre14}.

A caveat to this result is that the geometric distortion which is
corrected for by {\it astrodrizzle} necessitates resampling the pixels
in the image. We found, through trials using both the {\it multidrizzle} and
{\it astrodrizzle} packages, that the offset between the transformed
SN position and the position of the progenitor candidate was highly
sensitive to the choice of drizzling parameters applied to the
pre-explosion image (e.g. the subtraction of the sky background, the
shape of the drizzle kernel, the reduced pixel size or ``drop'' size
etc.). We note that this effect was not observed for brighter nearby 
stars, and appeared to arise solely due to the relative faintness of 
the candidate. In comparison with bright nearby surrounding stars, we 
found the position of the progenitor candidate could change by as much 
as $\sim 1.5$ pixels, due to the way in which flux associated with the 
candidate was allocated into the pixels in its locality.

We hence performed a second alignment between the pre-
and post-explosions, under the hypothesis that the {\texttt \_crj}
image was the least biased realisation of the detected progenitor
flux. We calculated the geometric transformation between the distorted
{\it F555W} {\texttt \_crj} image (j90n02021\_crj.fits) and the
undistorted post-explosion WFC3 {\it F555W} image, drizzled to
0.025\arcsec~pix$^{-1}$. To account for the distortion in the ACS
frame, a 4th order polynomial was used for the transformation, which
had an RMS error of 8 mas. The coordinates of iPTF13bvn were then
transformed to the {\texttt \_crj} image, where it lies at pixel
coordinates 2698.09,593.38. The position of the progenitor candidate
was measured using both the {\sc iraf} {\sc phot} package and with
{\sc dolphot} \citep{Dol00} to lie at 2698.0,593.83 and 2697.84,593.61
respectively. The progenitor candidate positions from {\sc phot} and
{\sc dolphot} are offset from the transformed SN position by
$\lesssim$~0.5 pixels ($\lesssim$~25 mas), however, they are also
offset from each other by 14 mas.

There are also significant differences found between the archival ACS
drizzled products provided in the STScI archives. There are two
products provided, the Hubble Legacy Archive (HLA) $\_$drz and the
MAST $\_$drc images which are both resampled from the original
detector pixels onto a grid of equal sky area pixels. The $\_$drc
files include CTE (charge transfer efficiency) corrections in the
pixel values, while the $\_$drz images do not. Photometry on $\_$drz
images thus requie CTE corrections after flux measurements. Using the
HLA $\_$drz images, an alignment between our WF3 drizzled frame
produces an RMS of 0.28 ACS pixels using 38 stars for alignment
(within the $geomap$ task of {\sc IRAF}). The positional uncertainty
between the SN and progenitor position is 0.73$\pm$0.42 ACS pixels
(where the error is the combined uncertainty in the alignment,
progenitor position and SN position). However using the same method
with the $\_$drc frame results in the positions matching to within the
uncertainty of 0.42 ACS pixels. In summary, while the position of the
SN and progenitor vary at the 1.5$\sigma$ level depending on which
drizzled product to use we conclude that they are likely coincident
within the errors based on our own manual $astrodrizzle$ ACS product
and the $\_$drc images.  The two papers published so far which have
discussed the progenitor identification and alignment (Cao et
al. 2013, Fremling et al. 2014) are not specifically clear which data
products have been used but we agree with these papers in suggesting
this is a likely progenitor candidate object.
The true test of whether iPTF13bvn and the progenitor candidate are
coincident will be at late times when it will be possible to see if
the latter has truly disappeared.

We performed Point Spread Function (PSF)-fitting photometry of the
pre-explosion {\texttt \_crj} images using the {\sc dolphot} package
\citep{Dol00}\footnote{http://americano.dolphinsim.com/dolphot/} with
the ACS module. Bad pixels were masked using the data quality images,
before {\sc dolphot} was run with the recommended parameters for
ACS/WFC data. The progenitor candidate was detected in all three of
the ACS filters. Interestingly, if {\sc dolphot} is run on each of the
filters separately, the magnitudes returned for the progenitor
candidate are $\sim$0.2 mag fainter than if all three filter images
were input to {\sc dolphot} together. 
We measure magnitudes on individual images in the VEGAMAG system of
{\it F435W} = 25.81$\pm$0.06, {\it F555W} = 25.86$\pm$0.08, {\it
  F814W} = 25.77$\pm$0.10.

To check the output of {\sc dolphot}, we also performed photometry on
the pre-explosion images using {\sc daophot} within {\sc
  iraf}. Photometry was performed on the {\texttt \_drc} files at the
native ACS/WFC pixel scale of 0.05\arcsec~pix$^{-1}$. The {\texttt
  \_drc} images have been corrected for both the inherent geometric
distortion of ACS, and for losses due to Charge Transfer Efficiency
(CTE). For each filter, a Point Spread Function (PSF) was constructed
from bright, isolated point sources. The modelled PSF was then fit
simultaneously to both the progenitor candidate and all surrounding
sources which may contribute flux at the position of the SN. The fit
was made within a small (2 pixel) radius centred on each source, and
the measured fluxes within this aperture were corrected to an infinite
aperture using the tabulated corrections in \cite{Sir05}. Finally, the
flux was converted to a magnitude in the {\it HST} VEGAMAG system
using the value of PHOTFLAM from the image header, and the flux of
Vega in the corresponding filter from the HST
webpages\footnote{http://www.stsci.edu/hst/acs/analysis/zeropoints/}.
We find magnitudes of {\it F435W}=25.79$\pm$0.10, {\it
  F555W}=25.73$\pm$0.07, {\it F814W}=25.99$\pm$0.22 for the progenitor
candidate, which agree favourably with the results of {\sc
  dolphot}. Because there is no clear reason to favour one over
the other we use a mean of the two values. This gives magnitudes for
the progenitor of {\it F435W} = 25.80$\pm$0.12, {\it F555W} =
25.80$\pm$0.11, {\it F814W} = 25.88$\pm$0.24.

\citet{Cao13} reported magnitudes for the progenitor of {\it
  F435W}=26.50$\pm$0.15, {\it F555W}=26.40$\pm$0.15 and {\it
  F814W}=26.10$\pm$0.20. It is not clear why there is such a great
difference between the two analyses. Other groups have also found
magnitudes that agree with those we derive (S. D. van Dyk,
priv. comm.). We can only suggest, in light of the fact we obtain
different results using {\sc dolphot} and {\sc daophot}, that the
results are dependent on the parameters given to these codes when the
photometry is derived and that any small error may be amplified.

The residual images after subtraction of the fitted PSFs were examined, and
do not show any gross over- or under-subtractions at the SN position.
However, it is clear that the background is not smooth at the SN position,
and late time observations after the SN has faded will be important to refine
the progenitor candidate photometry using template subtraction
\citep{Mau14}.

\begin{figure*}
\includegraphics[angle=0, width=180mm]{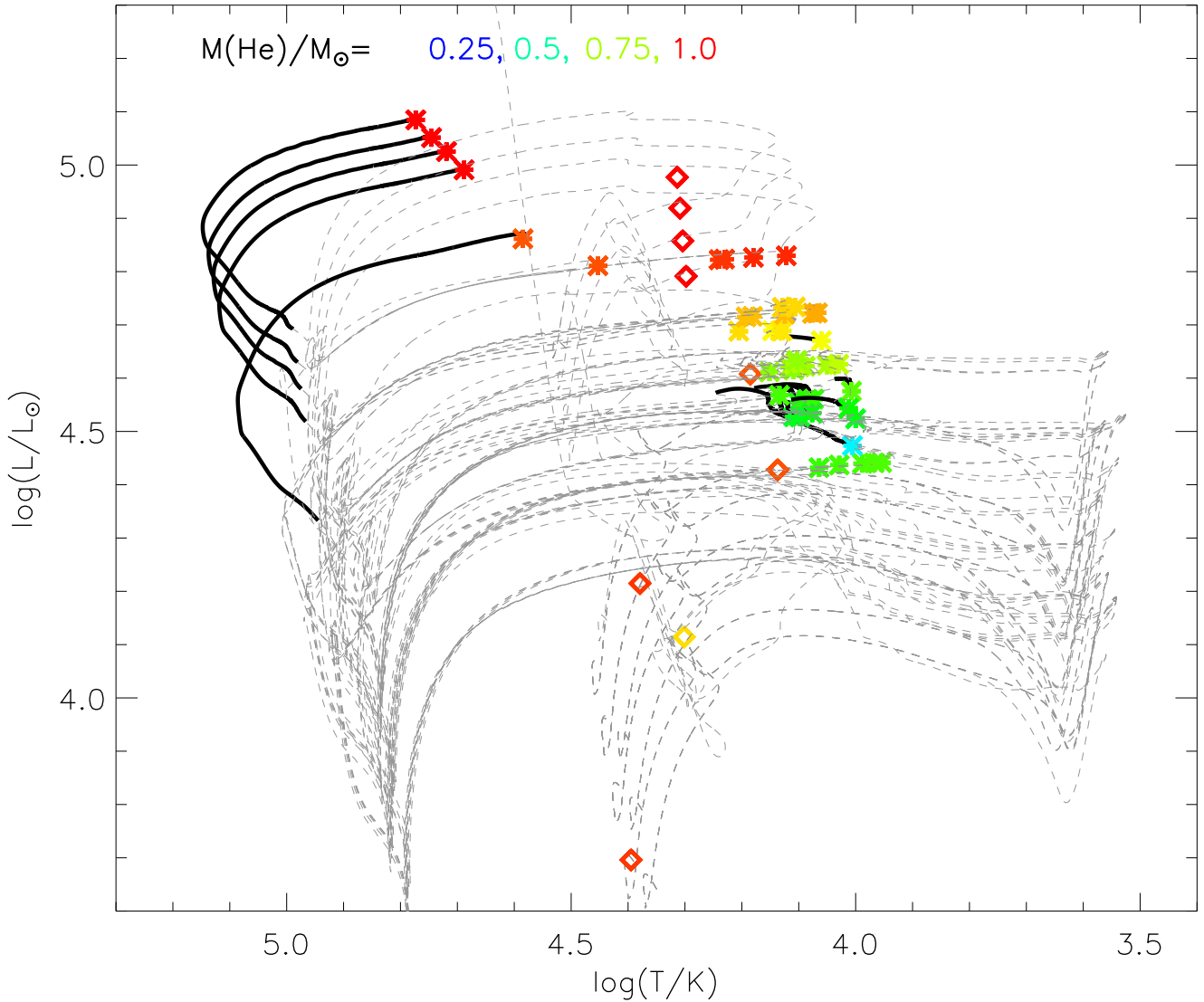}
\caption{HR diagram showing the evolutionary tracks for models which
  match the constraints on the progenitor candidate of iPTF13bvn. Thin
  grey dashed lines - evolution tracks with hydrogen, thick black
  lines - evolution without hydrogen. Asterisks - end point of
  progenitor models, diamonds - location of secondary star at
  explosion, they are included to indicate the general locations
  possible for the secondary star. Colour indicates helium mass in the
  primary at the end point of the model.}
\label{fig1}
\end{figure*}

\section{Numerical Method}

The construction of the stellar models used in this paper have been
described in detail in \cite{Eld08}. Here we use these models and
compare them to the progenitor candidate in a similar method to as in
\cite{Eld13}, but now compare the models to an actual detection rather
than upper limits on progenitor magnitude.  In summary the stellar
models follow single and binary stars at two metallicities, $Z=0.008$
and 0.020, that are close to the metallicity inferred for NGC5806 of
12+log~[O/H] = 8.5 from \cite{Sma09}. In the models the primary effect
of metallicity is to vary the mass-loss rates via stellar winds. The
evolutionary models are then matched to WR atmosphere models from the
Potsdam group \citep[e.g.][]{San12} to enable their magnitudes to be
calculated, as discussed in \cite{Eld09}. The grid of models covers
initial masses of the primary from 5 to 120$M_{\odot}$ with mass
ratios, $m_2/m_1$ between 0.1 to 0.9 and initial separations in
$\log(a/R_{\odot})$ from 1.0 to 4.0.

The major difference in our method here to that of \cite{Eld13} is
that our aim is to demonstrate that single star WR models are not the
only possible progenitor and interacting binaries can fit the observed
source and fit the other constraints available. Therefore we compare
the detected source to the end points of our models rather than
considering the whole evolutionary track closer to the time of
core-collapse. The latter is a more apt method to use when attempting
to estimate accurate parameters for the progenitor and take into
consideration uncertainties in the stellar evolution models
themselves. But until post-explosion images are available to more
tightly constrain the progenitor magnitudes, we consider only the
final model end points. These are typically after the end of
core-carbon burning and only a few years before core-collapse.

\begin{figure}
\includegraphics[angle=0, width=87mm]{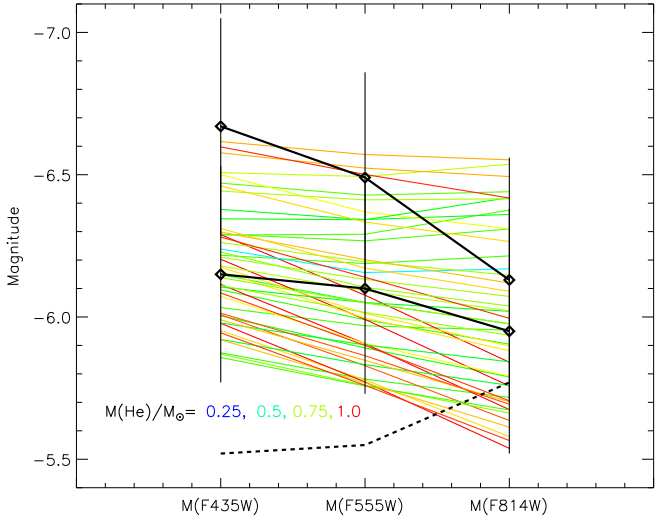}
\caption{SED of progenitor models compared to limits derived here with
  both the low and high extinction values used.  The error bars on the
  observed limits are mainly determined by the error in the distance
  to the host galaxy. Here the colours of the lines represent the
  helium abundance of the model as for the points in Figure 1. Most of
  the models are relatively cool with shallow or flat SEDs. }
\label{fig2}
\end{figure}

We have searched through our grid of models for stars which would give
rise to a hydrogen-free SN and compared the magnitude of these models
to the magnitude derived in Section \ref{s2}. With our assumed
distance the absolute magnitudes for the progenitors candidate are,
F435W=-5.96$\pm$0.38, F555W=-5.97$\pm$0.38 and
F814W=-5.88$\pm$0.43. We correct these magnitudes with the
\cite{Ber14} and \cite{Fre14} extinction values to obtain our final
magnitudes of between -6.15 to -6.67, -6.10 to -6.49 and -5.95 to -6.13
for the progenitor candidate. These comparisons were made in the {\it
  HST} filter system to avoid the additional uncertainty from
converting to the {\it UBVRI} system. The upper limit of possible
magnitudes are taken from magnitudes calculated with the higher
extinction value and the lower bound is from the lower extinction
value. We list the set of progenitor models which match the observed
magnitudes of the progenitor candidate within the error bars, and
within the range allowed by the uncertainty in extinction, in Table
\ref{tab1}. The evolutionary tracks of these models are plotted on a
Hertzsprung-Russell diagram in Figure \ref{fig1}, along with their
spectral energy distributions (SED) compared to the observed
magnitudes in Figure \ref{fig2}.  In most cases the SED is dominated
by the primary, apart from in the few cases where the final mass of
the secondary star is similar to the primary star's initial mass. We
caution however, that the effect of mass transfer can cause dramatic
evolutionary changes in the secondary star and much of the relevant
physics is uncertain, as discussed by \cite{Cla11}.

\begin{table*}
\caption{Physical parameters of the binary progenitor models which
  match the observed constraints on the progenitor of
  iPTF13bvn. Models where the primary mass has an asterisk beside it
  are systems with a compact objects for a secondary; the secondary
  masses of 0.6, 1.4 or 2.0$M_{\odot}$ correspond to a white dwarf,
  neutron star or black hole respectively. The given magnitudes are
  for the combined system of primary and secondary together. All
  systems however are dominated by the emission from the primary
  star. All masses, radii and luminosities are in given in units of $M_{\odot}$, $R_{\odot}$ and $L_{\odot}$ respectively. Surface temperatures are given in Kelvin.}
\label{tab1}
\begin{tabular}{ccccccccccccccccccccc}
\hline
\hline

 $M_{1,i}$ & $M_{2,i}$  & $\log(a/R_{\odot})$   &  $R_1$  & $\log T_1$   & $\log L_1$ &  $M_{1,f}$ &  $M_{2,f}$ & $M_{\rm H}$ & $M_{\rm He}$  &  $M_{\rm ej}$  &  $A*$  & $F435W$&  $F555W $&$F814W$ \\
\hline
\hline
\multicolumn{15}{c}{$Z=0.020$}\\
    9 &     8.1 &     2.25 &    41 &     4.06 &     4.43 &     2.05 &     8.4 &     0.00 &     0.66 &     0.60 &     0.35 &    -5.87 &    -5.78 &    -5.72 \\
    9 &     2.7 &     2.50 &    48 &     4.03 &     4.44 &     2.07 &     2.7 &     0.00 &     0.67 &     0.62 &     0.38 &    -6.03 &    -5.97 &    -5.95 \\
    9 &     4.5 &     2.50 &    59 &     3.99 &     4.44 &     2.08 &     4.6 &     0.00 &     0.68 &     0.63 &     0.41 &    -6.22 &    -6.19 &    -6.21 \\
    9 &     6.3 &     2.50 &    65 &     3.97 &     4.44 &     2.09 &     6.5 &     0.00 &     0.68 &     0.64 &     0.43 &    -6.29 &    -6.27 &    -6.31 \\
    9 &     8.1 &     2.50 &    69 &     3.95 &     4.44 &     2.09 &     8.4 &     0.00 &     0.68 &     0.64 &     0.44 &    -6.29 &    -6.29 &    -6.38 \\
   10 &     5 &     2.25 &    40 &     4.10 &     4.55 &     2.29 &     5.1 &     0.00 &     0.64 &     0.83 &     0.65 &    -6.00 &    -5.89 &    -5.79 \\
   10 &     7 &     2.25 &    43 &     4.08 &     4.56 &     2.30 &     7.2 &     0.00 &     0.65 &     0.84 &     0.69 &    -6.10 &    -5.99 &    -5.90 \\
   10 &     9 &     2.25 &    45 &     4.07 &     4.56 &     2.31 &     9.3 &     0.00 &     0.66 &     0.87 &     0.71 &    -6.15 &    -6.05 &    -5.98 \\
   10 &     5 &     2.50 &    63 &     4.01 &     4.58 &     2.36 &     5.1 &     0.00 &     0.69 &     0.92 &     0.88 &    -6.47 &    -6.43 &    -6.44 \\
   11 &     9.9 &     2.75 &    29 &     4.21 &     4.69 &     2.82 &    10 &     0.00 &     0.84 &     1.37 &     0.87 &    -5.95 &    -5.78 &    -5.58 \\
   13 &    11.7 &     1.25 &    61 &     4.00 &     4.52 &     2.19 &    22 &     0.00 &     0.60 &     0.74 &     0.71 &    -6.38 &    -6.34 &    -6.36 \\
   17 &    15.3 &     1.50 &     4.4 &     4.69 &     4.99 &     4.21 &    26 &     0.00 &     1.01 &     2.70 &     1.74 &    -5.98 &    -5.77 &    -5.54 \\
   18 &    16.2 &     1.50 &     4.0 &     4.72 &     5.03 &     4.45 &    28 &     0.00 &     1.03 &     2.98 &     2.03 &    -6.11 &    -5.91 &    -5.67 \\
   19 &    17.1 &     1.50 &     3.6 &     4.75 &     5.05 &     4.65 &    29 &     0.00 &     1.02 &     3.20 &     2.30 &    -6.20 &    -5.99 &    -5.76 \\
   20 &    14 &     1.25 &     6.1 &     4.58 &     4.86 &     3.50 &    29 &     0.00 &     0.94 &     2.02 &     0.99 &    -6.01 &    -5.83 &    -5.64 \\
   20 &    18 &     1.50 &     3.3 &     4.77 &     5.08 &     4.91 &    30 &     0.00 &     1.01 &     3.41 &     2.82 &    -6.29 &    -6.08 &    -5.84 \\
   10* &     5 &     2.25 &    56 &     4.01 &     4.47 &     2.06 &     2.2 &     0.00 &     0.42 &     0.62 &     0.56 &    -6.24 &    -6.16 &    -6.17 \\
   11* &     3.3 &     2.50 &    38 &     4.15 &     4.69 &     2.75 &     1.4 &     0.00 &     0.82 &     1.27 &     1.07 &    -6.07 &    -5.91 &    -5.79 \\
   11* &     5.5 &     2.50 &    55 &     4.06 &     4.67 &     2.74 &     2.0 &     0.00 &     0.81 &     1.29 &     1.15 &    -6.50 &    -6.37 &    -6.31 \\
   11* &     3.3 &     2.75 &    41 &     4.13 &     4.69 &     2.81 &     1.4 &     0.00 &     0.83 &     1.35 &     1.05 &    -6.18 &    -6.01 &    -5.90 \\
   11* &     5.5 &     2.75 &    40 &     4.13 &     4.69 &     2.80 &     2.0 &     0.00 &     0.83 &     1.35 &     1.05 &    -6.16 &    -6.00 &    -5.88 \\
\hline
\multicolumn{15}{c}{$Z=0.008$}\\
    9 &     6.3 &     2.25 &    37 &     4.11 &     4.53 &     2.14 &     6.5 &     0.00 &     0.60 &     0.69 &     0.58 &    -5.92 &    -5.83 &    -5.76 \\
    9 &     8.1 &     2.25 &    39 &     4.10 &     4.53 &     2.14 &     8.4 &     0.00 &     0.60 &     0.70 &     0.59 &    -5.98 &    -5.90 &    -5.84 \\
    9 &     2.7 &     2.50 &    43 &     4.08 &     4.53 &     2.16 &     2.7 &     0.00 &     0.62 &     0.72 &     0.64 &    -6.11 &    -6.05 &    -6.02 \\
    9 &     6.3 &     2.50 &    59 &     4.01 &     4.54 &     2.19 &     6.5 &     0.00 &     0.63 &     0.75 &     0.76 &    -6.35 &    -6.34 &    -6.42 \\
   10 &     3 &     2.25 &    34 &     4.15 &     4.61 &     2.49 &     3.0 &     0.00 &     0.72 &     1.02 &     0.72 &    -5.87 &    -5.76 &    -5.66 \\
   10 &     5 &     2.25 &    40 &     4.12 &     4.62 &     2.51 &     5.1 &     0.00 &     0.74 &     1.06 &     0.81 &    -6.11 &    -6.02 &    -5.94 \\
   10 &     7 &     2.25 &    43 &     4.10 &     4.62 &     2.52 &     7.1 &     0.00 &     0.74 &     1.07 &     0.83 &    -6.21 &    -6.13 &    -6.07 \\
   10 &     9 &     2.25 &    45 &     4.09 &     4.62 &     2.52 &     9.3 &     0.00 &     0.74 &     1.07 &     0.85 &    -6.26 &    -6.19 &    -6.15 \\
   10 &     3 &     2.50 &    54 &     4.05 &     4.62 &     2.55 &     3.0 &     0.00 &     0.75 &     1.10 &     0.92 &    -6.44 &    -6.41 &    -6.42 \\
   10 &     7 &     2.50 &    60 &     4.03 &     4.63 &     2.58 &     7.2 &     0.01 &     0.77 &     1.12 &     0.92 &    -6.51 &    -6.49 &    -6.54 \\
   11 &     9.9 &     1.25 &    35 &     4.13 &     4.57 &     2.29 &    18 &     0.00 &     0.67 &     0.84 &     0.63 &    -5.86 &    -5.76 &    -5.67 \\
   11 &     7.7 &     2.25 &    31 &     4.19 &     4.71 &     2.93 &     7.8 &     0.01 &     0.86 &     1.48 &     0.98 &    -5.92 &    -5.77 &    -5.61 \\
   11 &     9.9 &     2.25 &    33 &     4.18 &     4.72 &     2.93 &    10 &     0.01 &     0.86 &     1.49 &     1.01 &    -5.99 &    -5.85 &    -5.70 \\
   11 &     3.3 &     2.50 &    43 &     4.12 &     4.72 &     2.96 &     3.3 &     0.01 &     0.87 &     1.48 &     1.16 &    -6.30 &    -6.20 &    -6.12 \\
   11 &     5.5 &     2.50 &    54 &     4.08 &     4.72 &     2.97 &     5.5 &     0.01 &     0.87 &     1.53 &     1.30 &    -6.58 &    -6.52 &    -6.49 \\
   11 &     7.7 &     2.50 &    56 &     4.07 &     4.72 &     2.98 &     7.8 &     0.01 &     0.88 &     1.50 &     1.32 &    -6.62 &    -6.57 &    -6.55 \\
   12 &     8.4 &     1.25 &    41 &     4.11 &     4.62 &     2.49 &    18 &     0.00 &     0.72 &     1.05 &     0.82 &    -6.14 &    -6.05 &    -5.98 \\
   12 &    10.8 &     1.25 &    42 &     4.11 &     4.63 &     2.56 &    20 &     0.00 &     0.76 &     1.10 &     0.82 &    -6.18 &    -6.09 &    -6.02 \\
   12 &     8.4 &     2.50 &    29 &     4.24 &     4.82 &     3.38 &     8.5 &     0.01 &     0.96 &     1.93 &     1.83 &    -5.94 &    -5.76 &    -5.57 \\
   12 &    10.8 &     2.50 &    30 &     4.23 &     4.82 &     3.39 &    11 &     0.01 &     0.97 &     1.94 &     1.90 &    -6.08 &    -5.90 &    -5.71 \\
   12 &     3.6 &     2.75 &    38 &     4.18 &     4.83 &     3.42 &     3.6 &     0.01 &     0.97 &     1.96 &     2.17 &    -6.28 &    -6.14 &    -6.00 \\
   12 &     6 &     2.75 &    49 &     4.12 &     4.83 &     3.44 &     6.0 &     0.02 &     0.98 &     1.94 &     2.57 &    -6.60 &    -6.50 &    -6.42 \\
   40 &    12 &     1.25 &    11 &     4.45 &     4.81 &     3.31 &    42 &     0.01 &     0.94 &     1.87 &     1.00 &    -6.01 &    -5.86 &    -5.69 \\
   10* &     5 &     2.50 &    43 &     4.10 &     4.63 &     2.54 &     2.1&     0.00 &     0.73 &     1.10 &     0.89 &    -6.23 &    -6.10 &    -6.04 \\
   11* &     3.3 &     2.50 &    43 &     4.13 &     4.73 &     2.95 &     1.4 &     0.01 &     0.84 &     1.50 &     1.33 &    -6.31 &    -6.17 &    -6.09 \\
   11* &     5.5 &     2.50 &    48 &     4.11 &     4.73 &     2.94 &     2.0 &     0.01 &     0.84 &     1.50 &     1.43 &    -6.46 &    -6.33 &    -6.27 \\
\hline
\hline
\end{tabular}
\end{table*}

\section{Results}
\label{s4}

The large number of possible progenitor models means we need to
consider also the secondary constraints from \cite{Cao13},
\cite{Fre14} and \cite{Ber14}. We consider the constraint on the
mass-loss rate, ejecta mass and the requirement for sufficient helium
to produce a Type Ib SN. We do not use the radius constraint, because
as pointed out by \citeauthor{Ber14}, this constraint is not as firm
as first thought.

The constraint on the mass-loss rate from \cite{Cao13} needs to be
considered with care.  \citeauthor{Cao13} assume a wind velocity of
1000~${\rm km \,s^{-1}}$ to derive a mass-loss rate of approximately
$3\times 10^{-5}\, M_{\odot}\,{\rm yr^{-1}}$. This calculation is
strongly dependent on the wind velocity assumed. While 1000~${\rm km\,
  s^{-1}}$ is a typical WR wind speed, \cite{Eld06} found that WR wind
speeds evolve towards the end of a star's evolution and vary with
final mass. Therefore a more reliable constraint is to consider the
wind density, which is dependent on fewer assumptions. We use the
dimensionless wind parameter, $A_{*}$, where $A_{*} =({\dot{M}/10^{-5}
  M_{\odot}\,{\rm yr^{-1}}})/ (v_{\rm wind} /1000 {\rm km
  \,s^{-1}})$. Therefore values of the order unity are similar to
those from the typical Wolf-Rayet star.  From \citeauthor{Cao13},
iPTF13bvn has a value of $A_{*}=3$ therefore somewhat dense compared
to the typical Wolf-Rayet wind. The mass-loss rate and wind velocity
for our models is calculated from \cite{Nug00} as described in
\cite{Eld06}. We find lower mass models such as helium stars may have
weaker mass-loss rates, but they also have slower wind velocities. We
require that our models have an $A_*$ value between 0.3 and 30.0,
allowing for an order of magnitude error in the measured value and in
our calculation of the model values. Most of our models fall within
this range of observed wind parameter. Typically our model wind
parameters cover a range between 0.4 to 3. The measured value is
dependent on other physical assumptions so we do not regard this as a
significant disagreement.

The ejecta mass derived by \cite{Fre14} for iPTF1bvn is around
$1.94^{+0.50}_{-0.58}M_{\odot}$. This should be considered a lower
limit as there may always be additional helium that is transparent and
unobservable as described by \cite{transparent}. We estimate an ejecta
mass for our models by calculating the binding energy of the star and
using this to estimate how much mass would be ejected if $10^{51}$ergs
of energy was injected into the envelope as described in
\citet{E04}. Only in cases where the binding energy of the stellar
envelope is higher than this would there be material left to fall back
onto the central proto-neutron star. As our models have an initial
mass less than 20$M_{\odot}$ we find that a neutron star is always
produced so the ejecta mass is effectively the final mass minus
1.4M$_{\odot}$. For each model a corresponding minimum observed ejecta
mass can be estimated by subtracting the amount of helium in the
stellar model from the ejecta mass quotes in Table 1. We constrain our
model selection again by requiring the ejecta mass to be less than
$3.5M_{\odot}$. This upper limit is estimated by using the upper limit
from the error in the ejecta mass and upto 1M$_{\odot}$ of helium
being transparent \citep{transparent}. We find our model ejecta masses
are in reasonable agreement with the value of \cite{Fre14}, typically
lying between 1 and 2 \msun.

The minimum amount of helium which a star must retain to the point of
core-collapse if it is to produce the spectroscopic signature of a
Type Ib SN is still somewhat uncertain \citep{Des11}. In nearly all our
models there is greater than 0.5$M_{\odot}$ of helium in the ejecta,
likely to be enough to provide the required Type Ib SN spectrum. We
note that in some of the models there is a small amount of hydrogen
left on the surface of the star at the end of our models. Because our
models end at the end of carbon burning it is possible that this
hydrogen would subsequently be removed, in addition the mass-loss
rates of such stars are highest, and least certain at the end of their
lives, when they become helium giants.

In summary the possible binary progenitors we find for iPTF13bvn
mainly have masses between 10 and 20 \msun. A large number of possible
binary progenitors for iPTF13bvn will be survived by a visible, albeit
faint, stellar companion. There is also a subset of systems where the
companion will be a compact object, and undetectable at optical
wavelengths. We do not predict the magnitude for the secondary
companion, as the parameters of this star will be strongly affected by
the mass transfer process. Finally we note that we do find that some
very massive stars with initial masses of 80 \msun\ do match our
magnitude range. However these have very low amounts of helium, large
ejecta masses of around 5 to 7$M_{\odot}$ and the SED is dominated by
the binary companion. In this case the observed SED therefore represents
the binary companion not the progenitor itself.

\section{Discussion \& Caveats}

In contrast to the conclusion of \cite{Gro13}, we cannot find any
single-star models which match the SED of the progenitor candidate for
iPTF13bvn. This is largely due to our revised magnitudes being
brighter than previously reported, especially with the brighter {\it
  F435W} magnitudes. In addition to the other constraints such as the
total mass and mass of helium ejected. We note that our single-star
models do not include rotation so our analysis does not rule out a
single star solution completely. Therefore similar to \cite{Ber14}, we
conclude that iPTF13bvn most likely did not come from a non-rotating
single-star progenitor.

We caution that current uncertainties in stellar models could weaken
this result. For example the role of envelope inflation of WR stars,
an increase in their radius due to radiation pressure on the
iron-opacity peak, is still the subject of research and
debate. Therefore the single-star radii could be smaller or greater
than expected from models. In addition mass-loss rates of WR stars are
still to some extent uncertain so models may lose less mass during the
WR phrase and therefore contain more helium when they explode.

From our models we favour a binary progenitor for iPTF13bvn, most
likely a low-mass helium giant in a binary system. While such a helium
giant would have a radius larger that the limit of $<$~5 R$_\odot$
derived by \cite{Cao13}. \cite{Ber14} have suggested that for this
supernovae, as for SN 2011dh \citep{Ber12}, detailed modelling
demonstrates that the initial constraints on the progenitor radius are
not as stringent as first suggested.

We stress that all binary models represent a ``best-guess'' as to the
evolution of massive interacting binary stars. The largest uncertainty
remains the contribution from the binary companion of the progenitor
to the SED of the progenitor system. As discussed by \cite{Sta09} and
\cite{Cla11} the evolution of these stars post mass-transfer is
uncertain, and they may be cooler than normally expected for a
main-sequence star. Detection of any surviving companion in late-time
imaging of the SN site will provide an important constraint on the
binary scenario. Further more a spectrum of the star may reveal that
it is rapidly rotating because of the binary interactions
\citep{demink13}.

\section{Conclusions}

We have presented revised magnitudes for the source that is coincident
with the supernova iPTF13bvn. The {\it F435W} is the most
significantly brighter by 0.7 mags. This changes the shape of the
source SED and therefore has a strong influence on the resulting
possible objects that can match the progenitor source.

Using these new magnitudes and allowing for a range of extinctions
measured by different methods we find that it is possible to match the
source and other secondary constraints with binary models that had
initial masses between 10 to 20$M_{\odot}$. This overlaps with the
model suggested by \cite{Ber14}.

More massive models tend to not fit the source SED without a bright
companion star. Therefore if the source still exists when the SN fades
then the progenitor was a more massive Wolf-Rayet star rather than a
lower mass helium giant. However we suggest that the latter is highly
favoured in light of the ejecta mass estimates of \cite{Ber14} and
\cite{Fre14}. This is also in agreement with the prediction that helium
giants would be easier to identify as the progenitor of a type Ib/c SN by
\citet{yoon}.

It is only with late-time imaging that a deeper insight will be gained
into the progenitor. This has been demonstrated by analysis of even
the relatively well understood progenitors of Type IIP SNe
\citep{Mau14}.  If a surviving companion star is found at the site of
iPTF13bvn, then for the first time the binary evolution of a Type Ib
SN progenitor can be studied in detail.

\section{Acknowledgements}

JJE acknowledges support from the University of Auckland. This work
was partly supported by the European Union FP7 programme through ERC
grant number 320360. SJS acknowledge funding from the European
Research Council under the European Union's Seventh Framework
Programme (FP7/2007-2013)/ERC Grant agreement n$^{\rm o}$ [291222] and
STFC grants ST/I001123/1 and ST/L000709/1.  The research of JRM is
supported by a Royal Society Research Fellowship.



\label{lastpage}
\bsp

\end{document}